\documentstyle[12pt]{article}

\title{\bf A $5D$ non compact and non Ricci flat Kaluza-Klein Cosmology }
\author{F. Darabi\thanks{e-mail:
f.darabi@azaruniv.edu}\\{\small Department of Physics, Azarbaijan
University of Tarbiat Moallem, 53714-161, Tabriz, Iran .}\\
{\small Research Institute for Astronomy and Astrophysics of
Maragha, 55134-441, Maragha, Iran.} }
\begin{document}
\maketitle
\begin{abstract}
A model universe is proposed in the framework of 5-dimensional
noncompact Kaluza-Klein cosmology which is not Ricci flat. The $4D$
part as the Robertson-Walker metric is coupled to conventional
perfect fluid, and its extra-dimensional part is coupled to a dark
pressure through a scalar field. It is shown that neither early
inflation nor current acceleration of the 4$D$ universe would happen
if the non-vacuum states of the scalar field would contribute to
4$D$ cosmology.
\end{abstract}
\newpage
\section{Introduction}

According to the old suggestion of Kaluza and Klein the $5D$ vacuum
Kaluza-Klein equations can be reduced under certain conditions to
the $4D$ vacuum Einstein equations plus the $4D$ Maxwell equations.
Recently, the idea that our four dimensional universe might have
emerged from a higher dimensional space-time is receiving much
attention \cite{1}. One current interest is to find out in a more
general way how the $5D$ field equations relate to the $4D$ ones. In
this regard, a proposal was made recently by Wesson in that the $5D$
Einstein equations without sources $R_{AB}=0$ ( the Ricci flat
assumption ) may be reduced to the $4D$ ones with sources $G_{\alpha
\beta}=8 \pi G T_{\alpha \beta}$, provided an appropriate definition
is made for the energy-momentum tensor of matter in terms of the
extra part of the geometry \cite{Wesson}. Physically, the picture
behind this interpretation is that curvature in $(4+1)$ space
induces effective properties of matter in $(3+1)$ space-time. This
idea is known as {\it space-time-matter} or {\it modern
Kaluza-Klein} theory.

In a parallel way, the brane world scenario \cite{Brane} assumes
that our four-dimensional universe ( the brane ) is embedded in a
higher dimensional space-time ( the bulk ). The important ingredient
of the brane world scenario, unlike the space-time-matter theory, is
that the matter exists apart from geometry and is confined to the
brane, and the only communication between the brane and the bulk is
through gravitational interaction. The brane world picture relies on
a $Z_2$ symmetry and is inspired from string theory and its
extensions \cite{2}. This approach differs from the old Kaluza-Klein
idea in that the size of the extra dimensions could be large, more
or less similar to the idea in modern Kaluza-Klein theory.

On the other hand, the recent distance measurements of type Ia
supernova suggest an accelerating universe \cite{3}. This
accelerating expansion is generally believed to be driven by an
energy source which provides positive energy density and negative
pressure, such as a positive cosmological constant \cite{4}, or a
slowly evolving real scalar field called {\it quintessence}
\cite{5}. Since in a variety of inflationary models scalar fields
have been used in describing the transition from the
quasi-exponential expansion of the early universe to a power law
expansion, it is natural to try to understand the present
acceleration of the universe by constructing models where the matter
responsible for such behavior is also represented by a scalar field.
Such models are worked out, for example, in Ref \cite{6}. Bellini
{\it et al}, on the other hand, have published extensively on the
evolution of the universe from noncompact {\it vacuum} Kaluza-Klein
theory \cite{Bellini}. They used the essence of STM theory and
developed a 5D mechanism to explain, by a single scalar field, the
evolution of the universe including inflationary expansion and the
present day observed accelerated expansion.

In general, scalar fields are not the only possibility to describe
the current acceleration of the universe; there are (of course)
alternatives. In particular, one can try to do it by using some
perfect fluid but obeying "exotic" equations of state, the so-called
Chaplygin gas \cite{Chap}. This equation of state has recently
raised a certain interest because of its many interesting and, in
some sense, intriguingly unique features. For instance, the
Chaplygin gas represents a possible unification of dark matter and
dark energy, since its cosmological evolution is similar to an
initial dust like matter and a cosmological constant for late times
\cite{7}.

In this paper, motivated by higher dimensional theories, we are
interested in constructing a 5$D$ cosmological model which is not
Ricci flat, but is extended to be coupled to a higher dimensional
energy momentum tensor. This confronts the explicit idea of induced
matter in STM theory. Instead, we will show that the higher
dimensional sector of this model may induce a dark pressure, through
a scalar field, in four dimensional universe. The implications of
this dark pressure on the inflationary phase and current
acceleration of the universe will be discussed.

\section{The Model}

We start with the $5D$ line element
\begin{equation}
dS^2=g_{AB}dx^Adx^B, \label{0}
\end{equation}
in which $A$ and $B$ run over both the space-time coordinates
$\alpha, \beta$ and one non compact extra dimension indicated by
$4$. The space-time part of the metric $g_{\alpha \beta}=g_{\alpha
\beta}(x^{\alpha})$ is assumed to define the Robertson-Walker line
element
\begin{equation}
ds^2=dt^2-R^2(t)\left(\frac{dr^2}{(1-kr^2)}+r^2(d\theta^2+\sin^2\theta
d\phi^2 )\right), \label{1}
\end{equation}
where $k$ takes the values $+1, 0, -1$ according to a close, flat or
open universe, respectively. We also take the followings
$$
g_{4 \alpha}=0, \:\:\:\: g_{4 4}=\epsilon\Phi^2(x^{\alpha}),
$$
where $\epsilon^2=1$ and the signature of the higher dimensional
part of the metric is left general. This choice has been made
because any fully covariant $5D$ theory has five coordinate degrees
of freedom which can lead to considerable algebraic simplification,
without loss of generality. Unlike the noncompact vacuum
Kaluza-Klein theory, we will assume the fully covariant $5D$
non-vacuum Einstein equation
\begin{equation}
G_{AB}= 8 \pi G T_{AB}, \label{2}
\end{equation}
where $G_{AB}$ and $T_{AB}$ are the $5D$ Einstein tensor and
energy-momentum tensor, respectively. Note that the $5D$
gravitational constant has been fixed to be the same value as the
$4D$ one. In the following we use the geometric reduction from 5$D$
to 4$D$ as appeared in \cite{Ponce}. The $5D$ Ricci tensor is given
in terms of the $5D$ Christoffel symbols by
\begin{equation}
R_{AB}= \partial_C \Gamma^C_{AB}-\partial_B \Gamma^C_{AC}+
\Gamma^C_{AB}\Gamma^D_{CD}-\Gamma^C_{AD}\Gamma^D_{BC}. \label{3}
\end{equation}
The $4D$ part of the $5D$ quantity is obtained by putting $A
\rightarrow \alpha$, $B \rightarrow \beta$ in (\ref{3}) and
expanding the summed terms on the r.h.s by letting $C \rightarrow
\lambda, 4$ etc. Therefore, we have
\begin{equation}
\hat{R}_{\alpha \beta}=\partial_{\lambda} \Gamma^{\lambda}_{\alpha
\beta}+\partial_4 \Gamma^4_{\alpha \beta}-\partial_{\beta}
\Gamma^{\lambda}_{\alpha \lambda}-\partial_{\beta} \Gamma^4_{\alpha
4}+ \Gamma^{\lambda}_{\alpha \beta}\Gamma^{\mu}_{\lambda
\mu}+\Gamma^{\lambda}_{\alpha \beta}\Gamma^{4}_{\lambda
4}+\Gamma^{4}_{\alpha \beta}\Gamma^{D}_{4 D}-\Gamma^{\mu}_{\alpha
\lambda}\Gamma^{\lambda}_{\beta \mu}-\Gamma^{4}_{\alpha
\lambda}\Gamma^{\lambda}_{\beta 4}-\Gamma^{D}_{\alpha
4}\Gamma^{4}_{\beta D},\label{4}
\end{equation}
where $\hat{}$ denotes the $4D$ part of the $5D$ quantities. One
finds the $4D$ Ricci tensor as a part of this equation which may be
cast in the following form
\begin{equation}
\hat{R}_{\alpha \beta}={R}_{\alpha \beta}+\partial_4
\Gamma^4_{\alpha \beta}-\partial_{\beta} \Gamma^4_{\alpha 4}+
+\Gamma^{\lambda}_{\alpha \beta}\Gamma^{4}_{\lambda
4}+\Gamma^{4}_{\alpha \beta}\Gamma^{D}_{4 D}-\Gamma^{4}_{\alpha
\lambda}\Gamma^{\lambda}_{\beta 4}-\Gamma^{D}_{\alpha
4}\Gamma^{4}_{\beta D}.\label{5}
\end{equation}
Evaluating the Christoffel symbols for the metric $g_{AB}$ gives
\begin{equation}
\hat{R}_{\alpha \beta}={R}_{\alpha
\beta}-\frac{\nabla_{\alpha}\nabla_{\beta}\Phi}{\Phi}.\label{6}
\end{equation}
Putting $A=4, B=4$ and expanding with $C \rightarrow \lambda, 4$ in
Eq.(\ref{3}) we obtain
\begin{equation}
{R}_{4 4}=\partial_{\lambda} \Gamma^{\lambda}_{4 4}-\partial_{4}
\Gamma^{\lambda}_{4 \lambda}+ \Gamma^{\lambda}_{4
4}\Gamma^{\mu}_{\lambda \mu}+\Gamma^{4}_{4 4}\Gamma^{\mu}_{4
\mu}-\Gamma^{\lambda}_{4 \mu}\Gamma^{\mu}_{4 \lambda}-\Gamma^{4}_{4
\mu}\Gamma^{\mu}_{4 4}.\label{7}
\end{equation}
Evaluating the corresponding Christoffel symbols in Eq.(\ref{7})
leads to
\begin{equation}
{R}_{4 4}=-\epsilon\Phi \Box \Phi. \label{8}
\end{equation}
We now construct the space-time components of the Einstein tensor
$$
G_{AB}=R_{AB}-\frac{1}{2}g_{AB}R_{(5)}.
$$
In so doing, we first obtain the $5D$ Ricci scalar $R_{(5)}$ as
$$
R_{(5)}=g^{AB}R_{AB}= \hat{g}^{\alpha \beta}
\hat{R}_{\alpha\beta}+ g^{44}R_{44}= g^{\alpha \beta}(R_{\alpha
\beta}-\frac{\nabla_{\alpha}\nabla_{\beta}\Phi}{\Phi})+\frac{\epsilon}{\Phi^2}(-\epsilon\Phi
\Box \Phi)
$$
\begin{equation}
=R-\frac{2}{\Phi}\Box\Phi,\label{9}
\end{equation}
where the $\alpha 4$ terms vanish and $R$ is the $4D$ Ricci scalar.
The space-time components of the Einstein tensor is written
$\hat{G}_{\alpha \beta}=\hat{R}_{\alpha
\beta}-\frac{1}{2}\hat{g}_{\alpha \beta}R_{(5)}$. Substituting
$\hat{R}_{\alpha \beta}$ and $R_{(5)}$ into the space-time
components of the Einstein tensor gives
\begin{equation}
\hat{G}_{\alpha \beta}={G}_{\alpha \beta}+\frac{1}{\Phi}(g_{\alpha
\beta} \Box \Phi- \nabla_{\alpha}\nabla_{\beta}\Phi). \label{10}
\end{equation}
In the same way, the 4-4 component is written ${G}_{4 4 }={R}_{4 4
}-\frac{1}{2}g_{4 4}R_{(5)}$, and substituting ${R}_{4 4}$,
$R_{(5)}$ into this component of the Einstein tensor gives
\begin{equation}
G_{4 4}=-\frac{1}{2}\epsilon R\Phi^2, \label{11}
\end{equation}
We now consider the $5D$ energy-momentum tensor. The form of
energy-momentum tensor is dictated by Einstein's equations and by
the symmetries of the metric (\ref{1}). Therefore, we may assume a
perfect fluid with nonvanishing elements
\begin{equation}
{T}_{\alpha \beta}=(\rho+p){u}_{\alpha} {u}_{\beta}-p{g}_{\alpha
\beta}, \label{12}
\end{equation}
\begin{equation}
{T}_{44}=-\bar{p}g_{44}= -\epsilon\bar{p}\Phi^2, \label{14}
\end{equation}
where $\rho$ and $p$ are the conventional density and pressure of
perfect fluid in the $4D$ standard cosmology and $\bar{p}$ acts as a
pressure living along the higher dimensional sector. Hence, the
field equations (\ref{2}) are to be viewed as {\it constraints} on
the simultaneous geometric and physical choices of $G_{AB}$ and
$T_{AB}$ components, respectively.

Substituting the energy-momentum components (\ref{12}), (\ref{14})
in front of the $4D$ and extra dimensional part of Einstein tensors
(\ref{10}) and (\ref{11}), respectively, we obtain the field
equations\footnote{The $\alpha 4$ components of Einstein equation
(\ref{2}) result in
$$
{R}_{\alpha 4}=0,
$$
which is an identity with no useful information.}
\begin{equation}
G_{\alpha \beta}=8 \pi G [(\rho+p)u_{\alpha} u_{\beta}-pg_{\alpha
\beta}]+\frac{1}{\Phi}\left[\nabla_{\alpha}\nabla_{\beta}\Phi-\Box
\Phi g_{\alpha \beta}\right], \label{15}
\end{equation}
and
\begin{equation}
R=16 \pi G \bar{p}.\label{16}
\end{equation}
By evaluating the $g^{\alpha \beta}$ trace of Eq.(\ref{15}) and
combining with Eq.(\ref{16}) we obtain
\begin{equation}
\Box\Phi=\frac{1}{3}(8\pi G(\rho-3p)+16 \pi G \bar{p})\Phi
.\label{18}
\end{equation}
This equation infers the following scalar field potential
\begin{equation}
V(\Phi)=-\frac{1}{6}(8\pi G(\rho-3p)+16 \pi G \bar{p})\Phi^2,
\end{equation}
whose minimum occurs at $\Phi=0$, for which the equations (\ref{15})
reduce to describe a usual $4D$ FRW universe filled with ordinary
matter $\rho$ and $p$. In other words, our conventional $4D$
universe corresponds to the vacuum state of the scalar field $\Phi$.
From Eq.(\ref{18}), one may infer the following replacements for a
nonvanishing $\Phi$
\begin{equation}
\frac{1}{\Phi}\Box \Phi = \frac{1}{3}(8\pi G(\rho-3p)+16 \pi G
\bar{p}),\label{19}
\end{equation}
\begin{equation}
\frac{1}{\Phi}\nabla_{\alpha}\nabla_{\beta}\Phi = \frac{1}{3}(8\pi
G(\rho-3p)+16 \pi G \bar{p})u_{\alpha}u_{\beta}.\label{20}
\end{equation}
Putting the above replacements into Eq.(\ref{15}) leads to
\begin{equation}
G_{\alpha \beta}=8 \pi G [(\rho+\tilde{p})u_{\alpha}
u_{\beta}-\tilde{p}g_{\alpha \beta}], \label{22}
\end{equation}
where
\begin{equation}
\tilde{p}=\frac{1}{3}(\rho+2\bar{p}).\label{23}
\end{equation}
This energy-momentum tensor effectively describes a perfect fluid
with density $\rho$ and pressure $\tilde{p}$. It is very interesting
that the contributions of the non-vacuum states of the scalar field
at higher dimension cancels out exactly the physics of pressure $p$
in four dimensions. The field equations lead to two independent
equations
\begin{equation}
3\frac{\dot{a}^2+k}{a^2}=8 \pi G \rho, \label{24}
\end{equation}
\begin{equation}
\frac{2a\ddot{a}+\dot{a}^2+k}{a^2}=-8 \pi G \tilde{p}. \label{25}
\end{equation}
Differentiating (\ref{24}) and combining with (\ref{25}) we obtain
the conservation equation
\begin{equation}
\frac{d}{dt}(\rho a^3)+\tilde{p}\frac{d}{dt}(a^3)=0. \label{26}
\end{equation}
The equations (\ref{24}) and (\ref{26}) can be used to derive the
acceleration equation
\begin{equation}
\frac{\ddot{a}}{a}=-\frac{4 \pi G}{3}(\rho+3\tilde{p})=-\frac{8 \pi
G}{3}(\rho+\bar{p}). \label{27}
\end{equation}
If we choose the open universe ($k=-1$) in Robertson-Walker metric
(\ref{1}) so that $R=-a^{-2}$ and $\bar{p}= -\frac{1}{16\pi
G}a^{-2}$,  and insert a power law behavior $\rho=Aa^{\alpha}$ into
the conservation equation (\ref{26}), then we obtain\footnote{A
close universe $k=1$, $R>0$ will result in $\rho=-\frac{1}{16\pi
G}a^{-2}<0$ which is not physically viable.}
\begin{equation}
\rho=\frac{1}{16\pi G}a^{-2}>0.\label{28}
\end{equation}
By substituting $\rho$ and $\bar{p}$ into the acceleration equation
(\ref{27}) we find
\begin{equation}
\ddot{a}=0.
\end{equation}
Therefore, we conclude that the contributions of non-vacuum states
of the scalar field, living along the higher dimension, can lead to
zero acceleration of the 4$D$ universe, no matter which equation of
state $p=p(\rho)$ is used.

\section*{Conclusion}

In this paper, we have studied a $(4+1)$-dimensional metric subject
to a $(4+1)$ dimensional energy-momentum tensor in the framework of
noncompact Kaluza-Klein theory. The $4D$ part of the metric is taken
to be Robertson-Walker one subject to the conventional perfect fluid
with density $\rho$ and pressure $p$, and the extra-dimensional part
endowed by a scalar field is subject to the dark pressure $\bar{p}$.
By writing down the reduced $4D$ and extra-dimensional components of
$5D$ Einstein equations we found that our 4$D$ universe corresponds
to the vacuum state of the scalar field. It turned out that the
contributions of the non-vacuum states of the scalar field to the
4$D$ cosmology cancels out exactly the physics of pressure $p$ in
four dimensions and leads to zero acceleration of the 4$D$ universe
for any equation of state. In other words, if the non-vacuum states
of the scalar field at higher dimension would contribute to the 4$D$
cosmology then we would not see the current acceleration or even
expect early inflation of the universe. It is then possible to think
about other universes, living in excited states of the scalar field,
in which neither inflation nor acceleration ever happens.

This model, although introduced in the framework of noncompact
Kaluza-Klein theory, is not of Space-time-matter type as Bellini
{\it et al} have already worked out. So, a comparison between the
approach and results of this model and those of Bellini {\it et al}
is more constructive: In the Bellini {\it et al} approach, the Ricci
flat assumption $R_{AB}=0$ is made where matter as a whole is
induced by the {\it dynamics} of extra dimension. They developed a
5D mechanism to explain the ( neutral scalar field governed )
evolution of the universe from inflationary expansion towards a
decelerated expansion followed by the present day observed
accelerated expansion. In this model, however, we assumed a full
5$D$ Einstein equation coupled to a higher dimensional energy
momentum tensor whose components are all independent of 5th
dimension and its one extra dimensional component is a scalar field.
Reduction to four dimensions led us to 4$D$ Einstein equation
coupled to 4$D$ energy momentum tensor ( perfect fluid ) accompanied
by some terms of scalar field contribution induced from extra
dimension. Both models from different approaches try to address the
early inflation and current acceleration of the universe. Bellini
{\it et al} explain both early inflation and current acceleration of
the universe by a single scalar field. In the present model,
however, we show that the contributions of non-vacuum states of a
scalar field can destroy early inflation and current acceleration of
the universe. This result is independent of the signature $\epsilon$
by which the higher dimension takes part in the $5D$ metric.

Finally, we comment on the conceptual issue which is usually
considered in higher dimensional theories: {\it why we perceive the
$4$ dimensions of space-time and apparently do not see the fifth
dimension}? In old Kaluza-Klein theory this question is answered by
resorting to a cyclic condition imposed on the 5th coordinate. Brane
world cosmology also provides a mechanism by which matter and all
but gravitational interactions stick to the branes. In modern
Kaluza-Klein theory (STM), however, the matter itself and the
induced fifth force manifest as the direct results of the existence
of the 5th dimension. Similarly, in the present model we find that
the extra dimension may manifest through a dark pressure. However,
as we discussed above the existence of this dark pressure, through
non-vacuum states of the scalar field, will contradict the observed
acceleration and even early inflation of the universe, So, it turns
out that there is no such influence of dark pressure in our universe
and the reason why we do not see the higher dimension in this model
is that we are living in the vacuum state of the scalar field.

\section*{Acknowledgment}

This work has been supported by Research Institute for Astronomy and
Astrophysics of Maragha.

\end{document}